\documentclass[aps,reprint,prl,superscriptaddress,floatfix]{revtex4-2}

\usepackage{graphicx}

\begin{document}

\title{Strongly nonlinear antiferromagnetic dynamics in high magnetic fields}

\author{Pavel Stremoukhov}
\affiliation{FELIX Laboratory, Radboud University, Toernooiveld 7, 6525 ED Nijmegen, The Netherlands}

\author{Ansar Safin}
\affiliation{Kotel’nikov Institute of Radioengineering and Electronics, Russian Academy of Sciences, 125009, Moscow, Russia}
\affiliation{National Research University “Moscow Power Engineering Institute”, 111250, Moscow, Russia}

\author{Casper F. Schippers} 
\author{Reinoud Lavrijsen} 
\affiliation{Department of Applied Physics, Eindhoven University of Technology, P.O. Box 513, 5600 MB Eindhoven, Netherlands}

\author{Maurice Bal}
\author{Uli Zeitler}
\affiliation{High Field Magnet Laboratory (HFML – EMFL), Radboud University, Toernooiveld 7, 6525 ED Nijmegen, The Netherlands}

\author{Alexandr Sadovnikov}
\affiliation{Kotel’nikov Institute of Radioengineering and Electronics, Russian Academy of Sciences, 125009, Moscow, Russia}
\affiliation{Saratov State University, 410071, Saratov, Russia}

\author{Kamyar Saeedi Ilkhchy}
\affiliation{FELIX Laboratory, Radboud University, Toernooiveld 7, 6525 ED Nijmegen, The Netherlands}

\author{Sergey Nikitov}
\affiliation{Kotel’nikov Institute of Radioengineering and Electronics, Russian Academy of Sciences, 125009, Moscow, Russia}
\affiliation{Saratov State University, 410071, Saratov, Russia}
\affiliation{Moscow Institute of Physics and Technology, Dolgoprudny, 141701, Moscow Region, Russia}

\author{Andrei Kirilyuk}
\affiliation{FELIX Laboratory, Radboud University, Toernooiveld 7, 6525 ED Nijmegen, The Netherlands}

\begin{abstract}
Antiferromagnetic (AFM) materials possess a well-recognized potential for ultrafast data processing thanks to their intrinsic ultrafast spin dynamics, absence of stray fields, and  large spin transport effects. The very same properties, however, make their manipulation difficult, requiring frequencies in THz range and magnetic fields of tens of Teslas. Switching of AFM order implies going into the nonlinear regime, a largely unexplored territory. Here we use THz light from a free electron laser to drive antiferromagnetic NiO into a highly nonlinear regime and steer it out of nonlinearity with magnetic field from a 33-Tesla Bitter magnet. This demonstration of large-amplitude dynamics represents a crucial step towards ultrafast resonant switching of AFM order.
\end{abstract}

\maketitle

The dynamics of nonlinear systems has become a very active field of research in various branches of science \cite{benedek2012statics}. In condensed matter physics, the nonlinear behavior reveals itself in multiple  aspects, from phase transitions and shifting resonance frequencies, to solitons and chaos. In particular, in the case of magnetic materials, nonlinear dynamics is essential for the operation of spintronic devices whenever magnetic state switching via large amplitude precession is involved. It is well-known that, in ferromagnets, spin-wave instabilities destroy the homogeneous precession at high amplitudes \cite{suhl1957theory} and lead to chaotic magnetization dynamics \cite{rezende1990spin}. Striking example of this is the irreproducibility of ultrafast precessional magnetization reversal at picosecond time scales \cite{tudosa2004ultimate, kashuba2006domain}. 

While the fundamental understanding of magnetization dynamics resulted in vast applications of ferromagnetic materials \cite{Jubelium}, antiferromagnetic (AFM) materials for a long time were of academic interest only.
Only recently the fundamentally different and appealing features of AFMs started to attract wide attention \cite{vzelezny2018spin, vsmejkal2018topological, nvemec2018antiferromagnetic}. These are the absence of net magnetization and stray fields eliminating cross-talk in dense arrays and making AFM materials extraordinary stable. The fact that the strong exchange interaction is involved in even homogeneous precession of an AFM pattern, also unlike in ferromagnets, results in orders of magnitude faster spin dynamics in AFMs \cite{poots1961electrodynamics, vonsovskii1974magnetism, kimel2004laser}. However, it is the robustness of the AFM order that makes it far more difficult to control and to read out \cite{song2018manipulate}.
In particular, control of large-amplitude AFM dynamics remains a major challenge. In addition to the fundamental understanding of the enabling interaction mechanisms, strong THz fields are required to excite the AFM order into the nonlinear regime \cite{mukai2016nonlinear,Andreev1980}, while high magnetic fields must be applied to affect the equilibrium energy landscape. The combination of these two stimuli in a single setup could only be classified as an unmatched experimental tool. 

Here we report the results of strong excitation of AFM resonance in nickel oxide NiO as a function of applied magnetic fields. For the experiments, a combination of 33 Tesla Bitter magnet with high-intensity THz radiation from a free-electron laser was used. Interestingly, we observe a highly nonlinear dynamics of AFM spins, when the observed amplitude of precession does not depend on the excitation intensity. In addition, non-monotonous behavior of the spin-pumping into the adjacent Pt layer is observed, explained by the counter-acting frequency shifts resulting from the large amplitude and from the applied magnetic field. We believe that our results will be useful for the development of theoretical description of nonlinear AFM behavior, which may result in tunable and highly sensitive THz-frequency AFM devices such as detectors\cite{safin2020electrically}, emitters \cite{stremoukhov2019spintronic} and spectrum analyzers \cite{artemchuk2020}.

\begin{figure}[t]
        \centering
            \includegraphics[width=7cm]{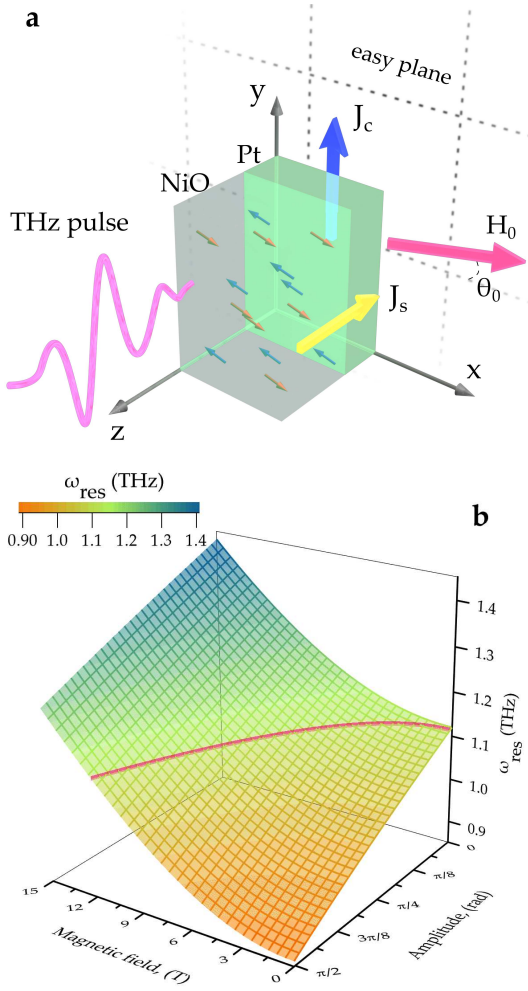}
    \caption{Excitation of antiferromagnetic resonance in NiO. (a) The experimental geometry, here \(J_{s}\) - spin current, \(J_{c}\) - charge current, \(H_{0}\) - the direction of externally applied magnetic field, \(\theta_{0}\) - an angle between the easy plane of the sample and \(H_{0}\). (b) Evolution of the AFMR frequency  as a function of amplitude of the steady state precession of the N\'eel vector and the applied magnetic field. The pink line is a contour line where AFMR frequency is 1.1 THz, as in small-amplitude approximation and in zero applied field.}
    \label{fig:AFMr}
    \end{figure}

Nickel oxide is the classical collinear type-II AFM, with approximately easy-plane magnetic anisotropy and Néel temperature of \(T_{N}\)=523 K \cite{kondoh1960antiferromagnetic}. In the paramagnetic phase above \(T_{N}\), NiO has the NaCl-type structure (point group m\(\overline{3}\)m). At temperatures below \(T_{N}\), in contrast, spins are coupled ferromagnetically within the \{111\} planes with neighboring planes being coupled antiferromagnetically by the  exchange interaction mediated by \(O^{2-}\) ions~\cite{roth1960antiferromagnetic}. In our experiments, a bi-layer structure consisting of an antiferromagnetic NiO layer and a heavy metal layer (Pt) is used, as sketched in Fig.\ \ref{fig:AFMr}a. Magnetic field is applied in the easy plane of NiO and thus also parallel to the NiO/Pt interface.

The magnetic structure of NiO is  described by two antiferromagnetic sublattices \(\mathbf{M_{1}}\) and \(\mathbf{M_{2}}\) shown in Fig.\ 1(a) as red and blue arrows correspondingly. The magnetization dynamics of an antiferromagnet under action of both DC \(\mathbf{H}_0\) and AC \(\mathbf{h}_{AC}(t)\) magnetic fields is usually described in terms of N\'{e}el  vector $\mathbf{l}=(\mathbf{M_{1}}-\mathbf{M_{2}})/ | \mathbf{M}_1 + \mathbf{M}_2 |$, by the so-called ``sigma-model''~\cite{Baryakhtar1979, Zvezdin1979, Andreev1980}:
\begin{eqnarray}\label{eq:neel}
 & & \mathbf{l} \times \left( \frac{d^2\mathbf{l}}{dt^2}+ \Gamma_{\mathrm{eff}} \frac{d\mathbf{l}}{dt} -2\gamma \left[\frac{d \mathbf{l}}{dt}\times \mathbf{H}_0 \right] + \frac{\partial W_{AFM}}{\partial \mathbf{l}}\right) \nonumber \\
 & & =  \left[\mathbf{l}\times \gamma\frac{d\mathbf{h}_{AC}}{dt}\right]\times\mathbf{l}.
\end{eqnarray}
Here \(\Gamma_{\mathrm{eff}}\) is the spectral linewidth of the AFM resonance at the zero DC magnetic field \cite{Cheng2014}, \(\gamma\) is the gyromagnetic ratio. Here the vector product \(d\mathbf{l}/dt\times\mathbf{H}_0\) is the gyroscopic torque~\cite{Andreev1980} and \(W_{AFM}(\mathbf{l},\mathbf{H}_0)\) is the magnetic energy density in the presence of the DC magnetic field, which can be expressed in the form (for more details see~\cite{Andreev1980,Baryakhtar1979}):
\begin{equation}\label{eq:enegry}
W_{AFM}(\mathbf{l}, \mathbf{H}_0) = W_{a}(\mathbf{l}) +\frac{\gamma^2}{2}\left( \mathbf{H}_0 \cdot \mathbf{l} \right)^2,
\end{equation}
where \(W_{a}(\mathbf{l})\) describes the effective anisotropy energy and determines the \textit{nonlinearity} of the system. The resonance frequencies of the NiO crystal in the absence of magnetic field are determined by the exchange constant as well as the anisotropy constants and are equal to \(\omega_1~=~2\pi\cdot~0.2\)~THz and \(\omega_2~=~2\pi\cdot~1.1\)~THz~\cite{khymyn2017antiferromagnetic}. 

\begin{figure*}[t]
        \centering
            \includegraphics[width=15cm]{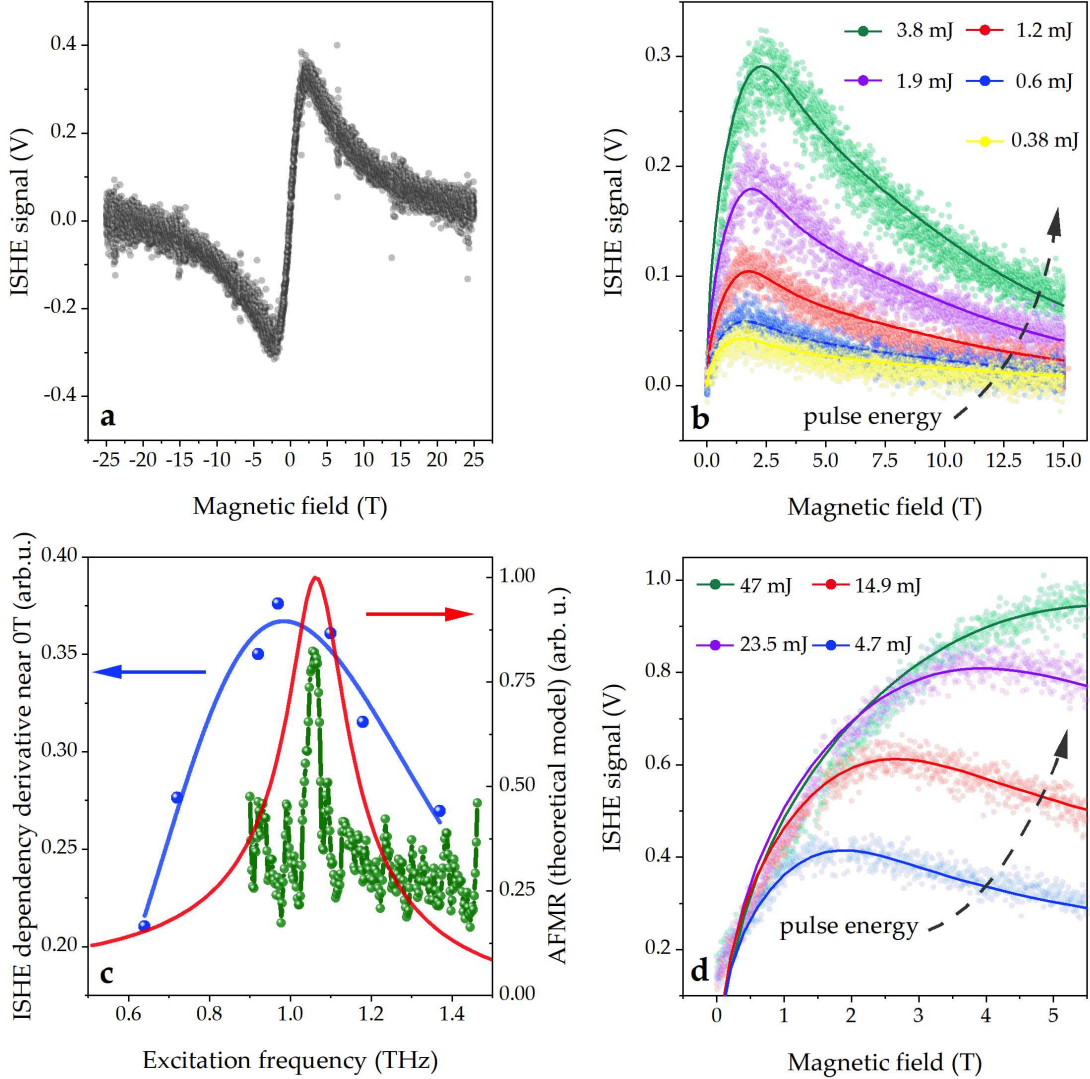}     
    \caption{Nonlinear behavior of high-amplitude antiferromagnetic resonance. (a) The antisymmetric shape of the iSHE voltage as a function of bias magnetic field at the excitation frequency of 1.1 THz. (b) Evolution of the detected voltage in magnetic field for various input light intensities and for the 1.1 THz excitation frequency. (c) Resonance behavior of the observed voltages: blue symbols -- derivative near 0 T field, red -- results of theoretical modeling, green -- results of Brillouin light scattering experiment. (d) Evolution of the detected voltage for the 0.72 THz excitation and in low magnetic fields.}
    \label{fig:ISHE}
\end{figure*}   

If one of the antiferromagnetic frequency modes of NiO \cite{doi:10.1063/1.5031213} is excited, a non-zero spin current can flow into the adjacent Pt layer. Cheng et al.\ developed a theoretical framework to understand dynamical spin injection from an AFM material undergoing coherent precession into an adjacent nonmagnetic material \cite{Cheng2014}. Contrary to the initial idea that spin pumping from antiparallel sublattice spins would cancel out, it was shown that coherent resonant rotations of different sublattice spins contribute constructively to the pumped spin current \cite{Cheng2014}.

Typically, in a uniaxial antiferromagnet and in zero applied magnetic field, the antiferromagnetic modes are degenerate \(\omega_1 = \omega_2\).  As the result of this degeneracy, spin pumping resulting from these two modes cancel. That makes spin current transport possible in uniaxial AFMs such as MnF\(_{2}\) and Cr\(_{2}\)O\(_{3}\) only in applied magnetic field  for  circular polarization of the incident electromagnetic wave \cite{delBarco}. However, NiO represents the class of easy-plane antiferromagnets. The presence of the additional, though weak, in-plane anisotropy in this material lifts the degeneracy \(\omega_1 \neq \omega_2\) and allows for magnonic spin current without external magnetic field \cite{rezende2016diffusive, Khymyn2017}. We observe however that this contribution is small compared to the field-induced signals, as can be seen in Fig.\ 2a where zero-field contribution is practically invisible as compared with the field-induced part. 

The DC component of the spin current density \(\mathbf{j}_{s}\) is described as \cite{tserkovnyak2002enhanced}
    \begin{equation}
     \mathbf{j}_s = \frac{\omega}{2\pi} \int_{0}^{2\pi/\omega} \frac{\hbar}{4\pi}g_{r}^{\uparrow\downarrow}\frac{1}{M_{s}^2} \Big[\mathbf{l}(t)\times \frac{d\mathbf{l}(t)}{dt}\Big]\,dt\, 
        \label{eq:spin_current}
    \end{equation}
where \(\omega\) and \(g_{\uparrow\downarrow}\) are the angular frequency of magnetization precession and the real part of the mixing conductance, respectively \cite{tokacc2015interfacial}. This shows that the excitation of dynamics of the N\'eel vector will result in a DC spin current \(\mathbf{j}_s\). The inverse spin-Hall effect in Pt layer converts this spin current into a DC voltage \cite{saitoh2006conversion, vlietstra2014simultaneous} due to the spin-dependent scattering \cite{van1963spin, PhysRevB.92.214403}, which is thus used in our experiments as the direct fingerprint of AFM dynamics.

 The  AFMR frequency can be found  by solving numerically the nonlinear equation (\ref{eq:neel}) taking into account Eq.\ (\ref{eq:enegry}) and magnetic field component of the AC electromagnetic field created by the THz source. The corresponding dependencies are shown in Fig.\ \ref{fig:AFMr}b. The pink line here serves as a guide to the eye that indicates the combination of magnetic field and precession amplitude that results in the same  value of the AFMR frequency.   The corresponding dependence of the frequency \(\omega_{res}\) of the resonant mode of NiO on the amplitude of stationary oscillations \(A_0\) can be approximated by the equation:
    \begin{equation}
     \omega_{res} = \sqrt{\omega_2^2 + \kappa \omega_{\text{H}}^2 - N\cdot A_0^2}.
        \label{eq:wres}
    \end{equation}
Here \(\kappa\) and \(N\) are phenomenological fit parameters that have been obtained by comparing experimental data with numerical simulations of model described by Eq.\ (\ref{eq:neel}) and \(\omega_{\text{H}}=\gamma\cdot H_0\). As can be seen from Eq.\ (\ref{eq:wres}), while the application of magnetic field increases the resonance frequency, the latter decreases with the amplitude of precession \(A_0\). The two effects can thus balance each other. 

 Detuning the AFMR frequency away from its resonant value (for NiO it is thus 1.1 THz at room temperature), still results in AFM precession of large amplitude, however smaller than in resonance. In particular, for experiments, this means that an AFMR excited in NiO further away from 1.1 THz will show a lower signal than an AFMR excited closer to 1.1 THz.  

Fig. \ref{fig:ISHE}a shows the dependence of the measured ISHE voltage as a function of the externally applied magnetic field, in the presence of radiation with central frequency 1.1 THz incident on the sample. First of all, the measured curve is antisymmetric with respect to the magnetic field, which happens because of the sign reversal of the spin current flowing into the Pt layer. This is in agreement with the behavior of the AFMR mode as a function of field \cite{delBarco} and time-reversal symmetry. However, strongly non-monotonic behavior of ISHE voltage as a function of magnetic field and excitation intensity, is much less obvious. 

To investigate this further, Figs.\ \ref{fig:ISHE}b,d show positive-field parts of this dependence measured for various intensities of THz radiation as well as for excitation frequencies on- and off resonance. Note that qualitatively the shape of signals is very similar in the case of resonant excitation (1.1 THz, Fig.\ \ref{fig:ISHE}b) as well as in the case of the off-resonant one (0.72 THz, Fig.\ \ref{fig:ISHE}d). Several features should be pointed out. First of all, note that in moderate applied magnetic fields and in high excitation intensity (see for example Fig.\ \ref{fig:ISHE}d, field values \(B\leq 1.5\)~T), the observed ISHE signal does not actually depend on the intensity. Thus, curves for 23.5 and 47 mJ pulse energy follow exactly the same trajectory up to above 2 T field. Also the dependence at lower pulse energy, 14.9 mJ, follows the same path until almost 1T field. This is a clear indication, that the observed excitation process is strongly nonlinear. On a side note, such behavior also directly rules out thermal effects such as thermally-induced spin-Seebeck effect \cite{PhysRevLett.115.266601,PhysRevLett.116.097204}, as the total thermal load on the sample is still low enough as to warrant linear heating regime.  

Second, after the initial linear increase of the ISHE voltage, signal reaches a maximum in the field range of 1--6~T, depending on the excitation intensity, followed by a slower decrease. Both the value of the signal at maximum as well as the magnetic field in which the maximum is achieved, depend on the THz radiation intensity. 

To prove that the observed ISHE signals are actually related to the expected AFMR, Fig.\ \ref{fig:ISHE}c shows the derivative of the signals in low fields $dI/dH$ as a function of the excitation frequency (blue dots).
The data clearly show a peak that is very similar to the one calculated theoretically. In addition, we show here the data of Brillouin light scattering experiment on the very same NiO film showing thermal magnons with $k\approx 0$ (for details, see Supplementary Information). The overlap is obvious, although the resonance measured with these high excitation levels is clearly broadened as well as shifted to low frequencies, as actually expected for large-amplitude behavior.  This indicates that similar to Ref.\ \cite{delBarco}, we detect spin currents induced by the excitation of AFMR in the antiferromagnetic layer. 

\begin{figure}[t]
    \centering
    \includegraphics[width=7.2cm]{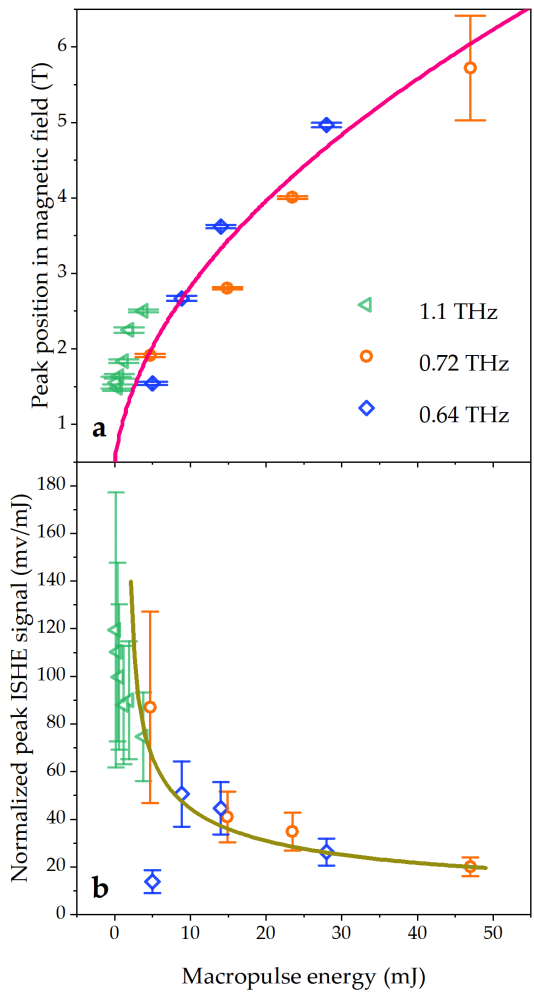}
    \caption{Features of non-monotonic ISHE behavior reflecting non-linear AFMR.  (a) AFMR peak position in bias magnetic field as a function of incident radiation power. Pink line is the same line that is shown in Fig.\ \ref{fig:AFMr}b. (b) Peak of ISHE signal normalized to the input pulse power in AFMR regime,  where dark yellow line represents result of mathematical modeling using Eq.\ (1)}
    \label{fig:fig.3}
\end{figure}

The observed non-trivial behavior of the ISHE signal as a function of the applied magnetic field can be understood as a direct consequence of strongly nonlinear excitation of AFMR. It derives from the combination of two factors that drive the system. First, the behavior of large-amplitude antiferromagnetic precession is essentially nonlinear in the sense that its frequency depend on the amplitude. As the amplitude grows, the frequency reduces first slowly, then faster, as illustrated in Fig.\ \ref{fig:AFMr}b. Therefore at some amplitude the frequency will be shifted out of the excitation bandwidth, and the precession saturates. Experimentally, we observe this saturation as an independence of the signal on the incident THz power in weak magnetic fields, see Fig.\ \ref{fig:ISHE}d. 

Applying magnetic field to this system has double effect. On the one hand, it creates a non-zero spin current into the adjacent Pt layer, allowing the detection of the antiferromagnetic precession via ISHE \cite{lebrun,delBarco}. Because of this, in low magnetic fields, we see an approximate proportionality of the observed ISHE signals to the applied field. On the other hand, the applied field also changes the AFMR frequency \cite{doi:10.1063/1.5031213}, see Fig.\ \ref{fig:AFMr}b. In strong fields, this change becomes sufficient to bring the system out of resonance. When this happens, the excitation becomes less efficient and results in a decrease of AFMR amplitude and also of the ISHE voltages. The field value at which this happens actually depends on the intensity of the excitation: the higher the intensity, the larger is the power broadening of the resonance line, and thus stronger field is required to move past the resonance condition. In experiment, we thus observe a shift of the maximum of the ISHE signal to higher fields (Figs.\ \ref{fig:ISHE}b,d), exactly in line with this expectation. The shape of the curves and the shift of the maximum are very well described by theory of nonlinear AFMR, which is proven by fitting using our model Eq.\ (\ref{eq:neel}).

The measurement were thus performed for a number of frequencies in the vicinity of NiO AFMR. Qualitatively, the same trend is observed for different values of the central frequency of incident THz light. However, the significant shift of local maximum in biasing magnetic field was registered. 

For further discussion of the nonlinear behavior of AFMR we show in Fig.\ref{fig:fig.3} the normalized height of the signal maximum as well as the field value where the maximum is achieved, as a function of macropulse energy. The data are compiled from the data sets obtained at different excitation frequencies. Surprisingly, on- and off-resonance  data roughly follow a single dependence, for both normalized maximum and the field where it is achieved. This can only be explained assuming that because of the very strong excitation, the resonance is pumped to the saturation from the very beginning. With that, the value of the AFMR frequency shifts according the Fig.\ \ref{fig:AFMr}b. The summary of the experimental data in Fig.\ 3a is reasonably fitted by the theoretical prediction of Eq.\ (\ref{eq:wres}). Note that the fitting line here is taken directly from Fig.\ \ref{fig:AFMr}b, where it shows the constant-frequency cross-section on the field-amplitude diagram. Thus interplay between the counteracting trends on the frequency introduced by the field and precession amplitude (see Fig.\ \ref{fig:AFMr}b) can only be weakly affected by the excitation power as the amplitude of the N\'eel vector precession nonlineary depends on the excitation amplitude. This is yet another convincing proof of the drastically nonlinear behavior of AFM spin dynamics in our experiments.

To conclude, we here demonstrate the realization of the first steps directed on the understanding  of ultrafast large-amplitude dynamics of antiferromagnets. The next crucial step would be a demonstration of a complete reversal of the AFM order driven by its resonance excitation. 

\begin{acknowledgments}
We gratefully acknowledge the Nederlandse Organisatie voor Wetenschappelijk Onderzoek (NWO-I) for their financial contribution, including the support of the FELIX Laboratory. This study was supported in part by the Russian Foundation for Basic Research (projects numbers 18-29-27020, 19-29-03015, 18-52-16006) and by the grant from the Government of the Russian Federation for state support of scientific research conducted under the guidance of leading scientists 
(project number 075-15-2019-1874). A part of the study was carried out within a state assignment for the Kotel'nikov Institute of Radio Engineering and Electronics of RAS. We also acknowledge support from the COST Action CA17123 ``Ultrafast opto-magneto-electronics for non-dissipative information technology'' (MAGNETOFON).
\end{acknowledgments}

\bibliography{references}

\newpage

\section*{Supplementary Information}

\subsection{Samples}

The NiO/Pt bi-layered structure was deposited on MgO(111) substrates with dimensions \(\approx\)4$\times$4~mm$^2$.
First, the substrates were cleaned by ion-beam milling to ensure high quality, crystalline interface. Without breaking the vacuum, layers of NiO and Pt were deposited. NiO was deposited by reactive magnetron sputtering from a pure Ni target in a 10:1 argon:oxygen mixture \((1 \times 10^{-3}\) mbar) at a temperature of 430\(^{\circ}\)C. Pt was grown by standard magnetron sputtering at room temperature in a pure argon environment \((1 \times 10^{-2}\) mbar). Afterwards, the samples were diced and wire-bonded to a chip-carrier. Over the course of the measurements, it became apparent that some considerations about the sample's geometry and area are essential for the proper investigation of spin-pumping from NiO to Pt. For these purposes, the samples were patterned in a Hall bar geometry \cite{standard1991f76,meyer1993methods}.

MgO was chosen as the substrate to grow the sample on due to reasonable transparency of the material for THz radiation in order to efficiently illuminate the NiO layer. Before doing so we performed TDS (time-domain spectroscopy) of the MgO (500 \(\mu m\)) sample described above. For this purpose TERA-K15-NL - Terahertz Time-Domain Spectrometer was employed. The transmissivity of the substrate is shown in Fig.\ \ref{fig:MgO} and is in a good agreement with the previous studies\cite{cunsolo1992refractive,han2008terahertz}, see below. From the plot, it follows that MgO substrate is reasonably transparent \((\approx 55\% - 60\%)\) near the area of the interest where AFM of NiO is expected to be observed \cite{doi:10.1063/1.5031213}.

\subsection{High magnetic field measurements}

\begin{figure}[ht]
    \centering
    \includegraphics[width=0.48\textwidth]{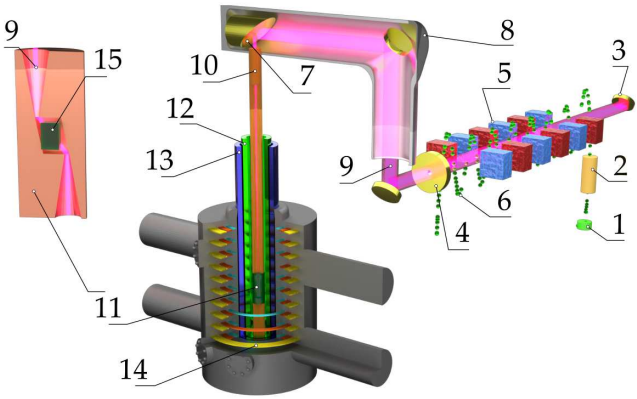}
    \caption{A schematic view of the experimental setup. The setup includes both the magnet at HFML and the FLARE free-electron laser. 1 - electron gun, 2 - accelerator, 3 - resonantor mirror, 4 - outcoupling resonator mirror, 5 - undulator magnets, 6 - electron bunches, 7 - focusing parabolic mirror, 8 - beamline, 9 - laser beam, 10 - sample holder stick/waveguide, 11 - sample area, 12 - liquid helium bath, 13 - liquid nitrogen bath, 14 - Bitter magnet coils, 15 - sample.}
    \label{fig:magsetup}
\end{figure}

The high magnetic field setup is shown in Fig.\ \ref{fig:magsetup}. The sample itself sits inside the insert, that is a cylindrical tube that works as an extension of the beamline inside the magnet and holds the sample. The insert is situated inside the continuous-flow cryostat that consist of a liquid helium bath (blue color) and a liquid nitrogen bath (green color).

The cryostat with the insert is placed inside a 33 Tesla Bitter magnet. The top of the insert is connected to a waveguide of the FLARE free-electron laser. The sample inside the insert is placed in the Voight geometry see Fig.\ \ref{fig:magsetup}. This is achieved by placing the chip holder with the sample inside a custom made brass insert with two 45$^{\circ}$ oriented polished brass surfaces. The surfaces serve as flat mirrors reflecting the light 90$^{\circ}$ and directing it through the sample. The external magnetic field is oriented in the sample plane. To the bottom of the insert, an InSb terahertz light detector (QMC Instruments) is connected in order to have a constant track of the light power that is transmitted through the sample. The whole setup is required to be in a vacuum because the region of interest \cite{doi:10.1063/1.5031213} includes a number of water vapor absorption lines for THz radiation. In the case of the particular experiment the sample, the insert, and the waveguide were evacuated using a number of pumps, the pressure inside the system was \(10^{-6}\) mBar. The signal path fully reproduces the one in the table-top stage of experiment.

Here we address some peculiarities of the setup that are important for experimental data presentation. Due to certain technological restrictions, direct measurement of the effective light power illuminating the sample becomes challenging. Another feature of the setup is that the free-electron-laser source for the THz region (FLARE) has nonconstant power distribution for a given frequency range (see attachments). However, there is still a possibility to measure stick input power in a place of beamline/cryostat coupling, see Fig.\ \ref{fig:magsetup}. The setup includes a set of attenuators in the beamline to control FLARE output macropulse energy.

\subsection{THz pulse time structure}

The time evolution of the THz radiation provided by a free-electron laser defines the detection scheme and the setup layout. In this case, the FLARE light source of the FELIX laboratory \cite{6665600} was used. The source covers the range from 0.24 to 3 THz (8 - 100 cm\(^{-1}\)).  The source has the following pulse structure: macropulses with a repetition rate of 5 Hz and duration 10 \(\mu\)s,  each macropulse is filled with micropulses with a repetition rate of 3 GHz.  The duration of each micropulse is defined by cavity detuning of the free-electron laser as well as . Further in this work, we treat the macropulse as a continuous pulse of THz light with a duration of 10 \(\mu\)s with an average macropulse energy of 100 mJ ignoring the micropulse structure.

\subsection{Model}

In Eq.\ (2) of the main text, the anisotropy energy of the "easy plane" biaxial NiO is modeled by the function:
\begin{equation}
W_{a}(\mathbf{l}) = -\frac{\omega_{\text{ex}}\omega_{\text{EA}}}{2} \left( \mathbf{l}\cdot \mathbf{e}_{\text{EA}} \right)^2 + \frac{\omega_{\text{ex}}\omega_{\text{HA}}}{2} \left( \mathbf{l}\cdot \mathbf{e}_{\text{HA}} \right)^2,
\end{equation}
where characteristic frequencies are defined as follows: \(\omega_{ex}~=~\gamma H_{ex}\), \(\omega_{EA}~=~\gamma H_{EA}\), \(\omega_{HA}=\gamma H_{HA}\), \(H_{ex}\) is the AFM internal exchange magnetic field, \(H_{e},H_{h}\) and \(\mathbf{e}_{EA} = \mathbf{x}\), \(\mathbf{e}_{HA} = \mathbf{z}\) are the AFM anisotropy easy and hard fields and unit vectors, respectively \cite{khymyn2017antiferromagnetic, Safin2022}. The corresponding values of all the constants listed in the paper are given in the table below.

The spectral linewidth of the AFM resonance \(\gamma_{eff}~=~\alpha_{eff}\omega_{ex}\) is determined by the effective damping \(\alpha_{eff}\) including Gilbert constant and spin-pumping terms ~\cite{Cheng2014}. Thus, the quality factor of the AFMR for the high frequency mode 1.1 THz and effective Gilbert damping \( \alpha_{eff} = 3.5\cdot 10^{-3}\) is equal to  \(Q=11.4\), which is typical for such AFM crystals \cite{vonsovskii1974magnetism}. 

The magnetic field component of the AC electromagnetic field $\mathbf{h}_{AC}=h_{AC}\mathbf{e_{\text{AC}}}\cdot \sin(\omega t + \phi_0)$ created by a laser source, where \(h_{AC}\), \(\omega\) and \(\phi_0\) are the amplitude, frequency and initial phase of the input signal, respectively. The \(\mathbf{e}_{AC}\) is the polarization vector of the electromagnetic wave, which in our case is oriented in the easy plane \(\mathbf{xy}\).

We rewrite the nonlinear Eq.\ (1) in a spherical coordinate system with in-plane polar angle \(\varphi(t)\) and azimuthal out-of-plane angle \(\theta(t)\) in a form:
\begin{equation}\label{eq:theta_phi}
\left\{
\begin{array}{c}
\begin{array}{c}
    \sin^2\theta\cdot
    \left(
    \begin{array}{c}
    \frac{d^2\varphi}{dt^2} + 
    \gamma_{eff}\frac{d\varphi}{dt} +\\ \frac{\sin2\varphi}{2}\bigl(\omega_{EA}\omega_{ex}+ 
    \omega_{y}^2-\omega_{x}^2\bigr)
    \end{array}
    \right) + \\[4pt]
    2\gamma(\mathbf{l}\cdot \mathbf{H}_0)\sin\theta\frac{d\theta}{dt} +  \omega_{x}\omega_{y}\sin^2\theta\cos2\varphi + \\[4pt] \sin2\theta\frac{d\varphi}{dt}\frac{d\theta}{dt} = \gamma (\mathbf{l}\cdot\frac{d\mathbf{h}_{AC}}{dt})\cos\theta,\\[4pt]
  \end{array}\\[20pt]
  \begin{array}{c}
    \frac{d^2\theta}{dt^2} +\gamma_{eff}\frac{d\theta}{dt} - \\[4pt] \frac{\sin2\theta}{2}
    \left(
    \begin{array}{c}
    \omega_{HA}\omega_{ex} + \omega_{EA}\omega_{ex}\cos^2\varphi-\\[4pt]
    - (\omega_{x}^2\cos^2\varphi+\omega_{y}^2\sin^2\varphi)
    \end{array}
    \right)- \\[4pt] 2\gamma\left(\mathbf{l}\cdot\mathbf{H}_0\right)\sin\theta\frac{d\varphi}{dt}+
    \omega_{x}\omega_{y}\frac{\sin2\theta}{2}\sin2\varphi = \\[4pt]
    \gamma\left(\frac{d h_{AC,x}}{dt}\sin\varphi - \frac{d h_{AC,y}}{dt}\cos\varphi
    \right),
  \end{array} 
\end{array}
\right.
\end{equation}
where \(\omega_{x,y}=\gamma(\left[\mathbf{x},\mathbf{y}\right]\cdot\mathbf{H}_0)\) and \(h_{AC,\left[x,y\right]}\) are the projections of a vector \(\mathbf{h}_{AC}\) into \(\mathbf{x,y}\), respectively. We numerically solved the system of non-linear equations (\ref{eq:theta_phi}) for different values of amplitude \(h_{AC}\) and frequency \(\omega\) of the EM wave. It was also solved for a number of different parameters using Powell's dog leg method, in particular its Python implementation.


The output ISHE voltage can be calculated as a result of finding the projection of the Eq.\ (2) on the $x$ axis and multiplying it by the phenomenological constant, which was selected  as a result of fitting the experimental data.

\subsection{THz-range transmission from TDS}

\begin{figure}[h!]
        \centering
            \includegraphics[width=0.48\textwidth]{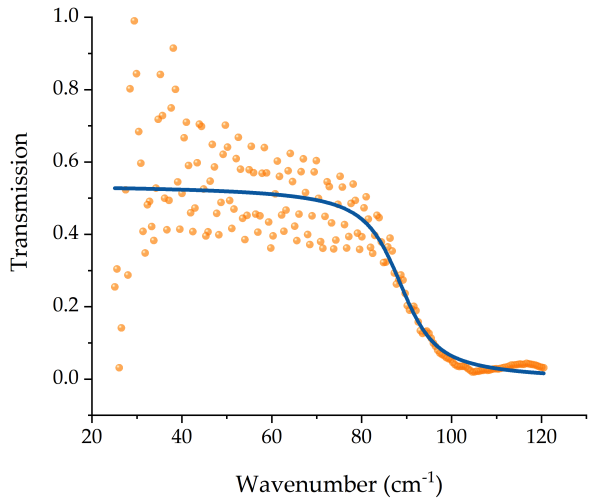}
            \caption{TDS characterization results for MgO(111) substrate crystal}
            \label{fig:MgO}
\end{figure}

The data from time-domain spectroscopy (TDS) are shown in Fig.\ \ref{fig:MgO} and confirm the transparency of the MgO substrate in the required frequency range. 

\subsection{Signals in the absence of magnetic field}

This experiment was carried out in two stages. The main experimental stage was carried out in the presence of a high magnetic field for the observation of non-linearly excited AFMR. There was however the preparation stage that was carried out in absence of a magnetic field. The goal of these measurements was to find, optimize and study the origin of the ISHE signal in the Pt layer.

\begin{figure}[h!]   
    \centering 
    \includegraphics[width=0.48\textwidth]{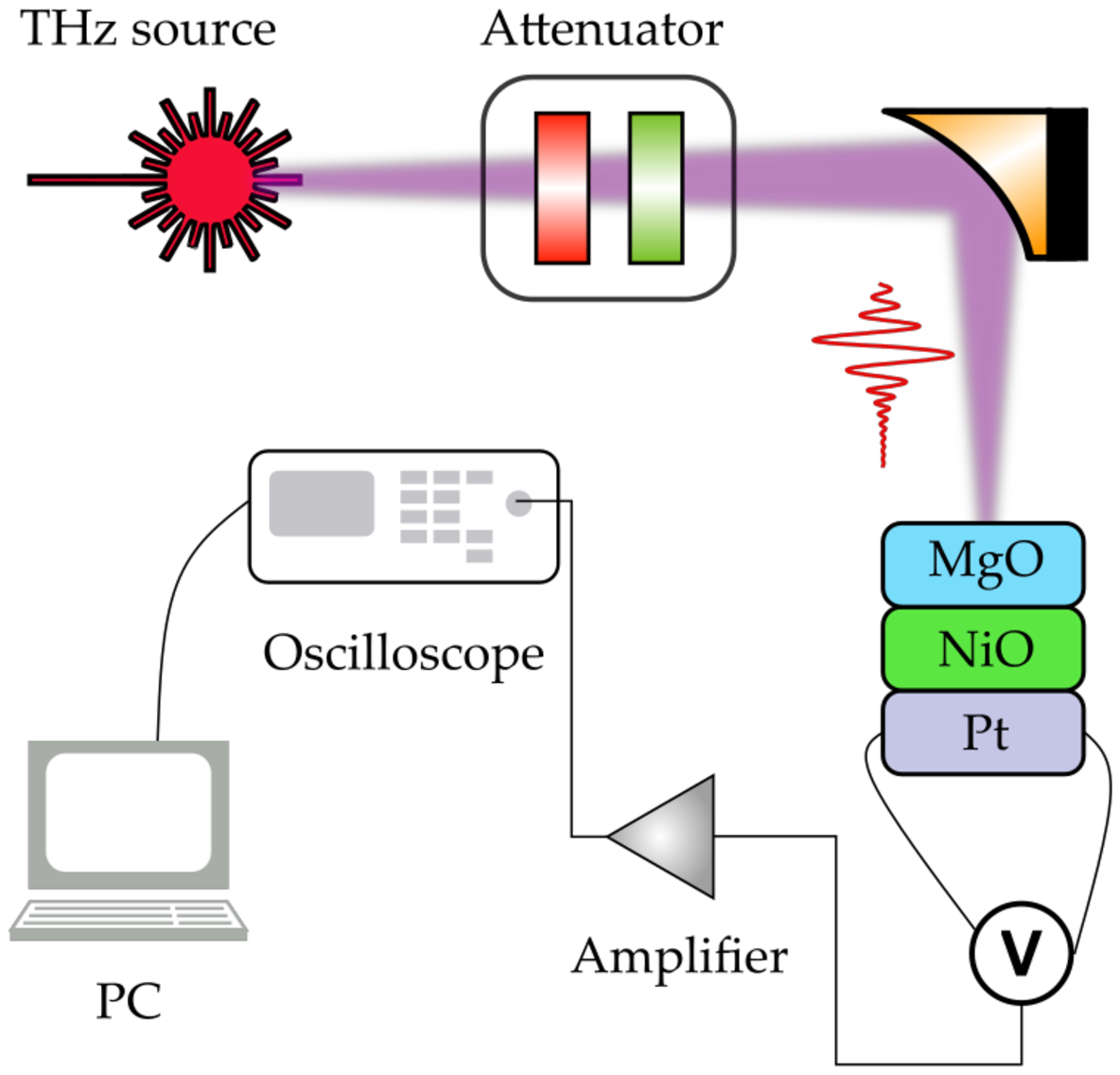}
    \caption{Table-top setup}%
    {{\small}}    
    \label{fig:tablesetup}
\end{figure}

The employed setup is depicted on Fig.\ \ref{fig:tablesetup}. The main goal of the setup is to expose the NiO layer of the sample to THz excitation and detect the resulting electric signals in Pt.

The THz radiation propagates through the optical attenuator consisting of a polarizer and motorized analyzer. This allows to smoothly control optical power according to Malus’s law\cite{damian2006malus}. The signal path is the following: the Pt layer is electrically wired to the chip holder, the chip holder is connected to SR560 Stanford Research Preamplifier with coaxial cable, amplified signal from the preamplifier goes to National Instruments PXIe-5162 digital oscilloscope. No external magnetic field was being applied at this part of the experiment. 

\begin{figure}[h!]
     \centering
         \includegraphics[width=0.45\textwidth]{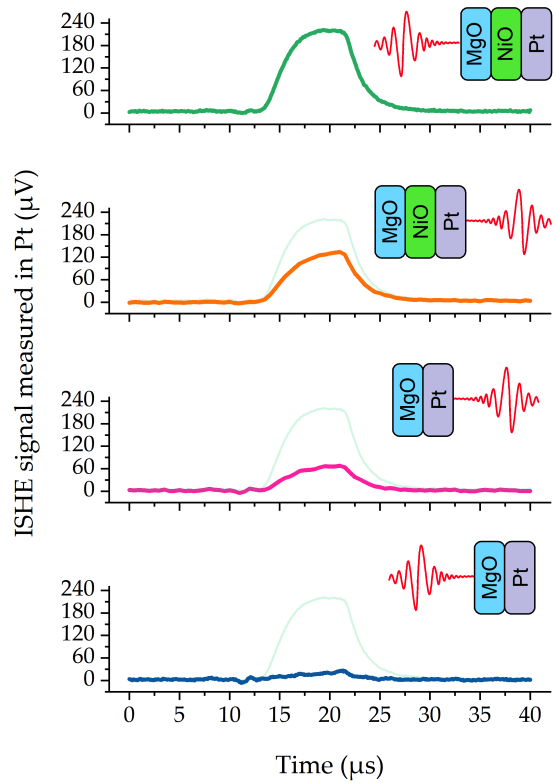}
         \label{fig:01a}
    \caption{Voltages measured from different sides of the sample.}
    \label{fig:ISHEsignal}
\end{figure}

One of the most important tasks of the experiment was to show the possibility of electrical detection of a spin-current by means of ISHE. However, it was important to make sure that the signal we have observed is not just a pure thermoelectric effect (Joule heating) as the sample is being exposed to a significant amount of optical power. To do so a series of tests were performed. As a reference, the following sample was used MgO(500\(\mu\)m)/Pt(10nm). The results were compared with the result obtained from the following sample MgO(500\(\mu\)m)/NiO(25nm)/Pt(10nm). Each sample was exposed to the light from the MgO side and the Pt side. The electrical signal was measured from the Pt side. The results can be seen in Fig.\ \ref{fig:ISHEsignal}.

 It would be reasonable to compare the green trace (Fig.\ \ref{fig:ISHEsignal}) with the blue trace to estimate the input of the Joule heating. From these plots, it is clear that the Joule heating contribution is about \(\approx 8\%\), points out the that \(\approx 92\%\) of the green trace are of the non-heating origin. That can be explained with two following scenarios. The first one is that the signal is a result of a non-resonant spin dynamics being excited in the magnetic lattice of NiO (due to the spin-Seebeck effect) that is being converted into electrical current in Pt via ISHE on the interface NiO/Pt. The second scenario is connected with a resonant dynamics being excited by THz pulse (due to antiferromagnetic resonance in NiO) that is being converted into electrical current in Pt via ISHE on the interface NiO/Pt.
 
 The pink trace (Fig.\ \ref{fig:ISHEsignal}), in this case, is explained with pure thermoelectrical effect (the Joule heating) in Pt. The effect is significantly more pronounced as compared to the blue trace that is explained by energy losses by absorption in the MgO together with back-reflection on the MgO/Pt interface in the case of the blue trace.
 
 The orange trace (Fig.\ \ref{fig:ISHEsignal}) is logically followed from the pink trace with additional contribution from a magnetic effect that originates in NiO. However, this contribution is smaller than in the case of the green term due to the energy losses on heating the Pt layer together with back reflection from Pt/air interface and back reflection from Pt/NiO interface.

\subsection{The fitting of experimental data}

For the experimental data fitting on Figs.\ 2(b,d) of the main text, the following parameters of the model (Eq.\ (\ref{eq:theta_phi})) were used:
\vfill

\begin{tabular}{ |p{1.8cm}|p{3cm}|p{3cm}|  }
\hline
\multicolumn{3}{|c|}{Model parameters list} \\
\hline
Parameter &values for Fig.\ 2(b) & values for Fig.\ 2(d) \\
\hline
\(\theta\)    &10\textdegree & 10\textdegree \\
\(\omega_{in}\) & 1.1 THz & 0.72 THz \\
\(\gamma_{eff}\) & 28.024 \(\frac{MHz}{T}\) & 28.024 \(\frac{MHz}{T}\)   \\
\(\omega_{HA}\) & 43.9 GHz & 43.9 GHz \\
\(\omega_{ex}\) & 27.5 THz & 27.5 THz\\
\(\alpha\) & \(3.4 \cdot 10^{-3}\) &\(1.1 \cdot 10^{-2}\) \\
\(h_{AC}\) &  & \\
\textit{green} & 0.2067 rad & 0.4921 rad \\
\textit{orange} & 0.1934 rad & 0.3107 rad \\
\textit{red} & 0.1741 rad & 0.2237 rad \\
\textit{blue} & 0.168 rad & 0.1777 rad \\
\textit{black} & 0.162 rad & \\
weighting & & \\
coefficients & & \\
\textit{green} & 0.6 & 0.46 \\
\textit{orange} & 0.45 & 0.565 \\
\textit{red} & 0.28 & 0.63 \\
\textit{blue} & 0.17 & 0.6 \\
\textit{black} & 0.135 & \\
\hline
\end{tabular}
\vfill

The main fitting parameter for the data is \(h_{AC}\) which is a direct representation of effective laser pulse energy. The \(h_{AC}\) is used to adjust the peak position of the dependencies on Figs.\ 2(b.d). The effective damping \(\alpha\) parameter was adjusted for the case of the 0.72 THz excitation. Possibly due to the high power of the laser light, additional non-linear mechanisms of effective damping were introduced. In case of the 1.1 THz excitation we kept \(\alpha\) close to theoretically predicted value \cite{khymyn2017antiferromagnetic}. Additionally to these parameters weighting coefficients were used to match theoretical curves with the experimental data in amplitude.

For Fig.\ 3(b) all the same parameter values as for Fig.\ 2(b) were used. The \(h_{AC}\) parameter is spanned in interval [0:1000].

For the Fig.\ 1(b) Eq.\ (4) was solved keeping all the parameter values the same, \(N \approx 1.28\cdot10^{25} rad/s\) and \(\kappa = 4.5\).

\subsection{Brillouin light scattering data}

\begin{figure}[h!]
     \centering
         \includegraphics[width=8cm]{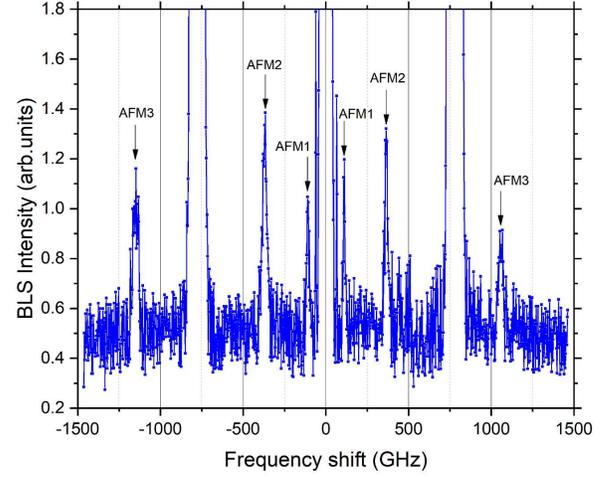}
         \label{fig:01}
    \caption{Backscattering Brillouin spectra of NiO taken in 2FSR operational mode. Identified AFM modes are indicated with AFM1, AFM2 and AFM3.}
    \label{fig:BLS}
\end{figure}

Brillouin light scattering (BLS) experiments were performed in the backscattering geometry using the Sandercock tandem interferometer in ``Two-free-spectral-ranges'' (2FSR) operation regime. This corresponds to the selected frequency range of inelastically scattered light from -1.5 THz to 1.5 THz. Fig.\ \ref{fig:BLS} demonstrates the BLS spectra in this regime and three well pronounced modes are indicated with AFM1, AFM2 and AFM3. In the resulting spectrum, three peaks in the Stokes and three peaks in the anti-Stokes parts can be distinguished, having frequencies of 110 GHz, 366 GHz and 1140 GHz, respectively. Peak AFM2 has the center frequency 366 GHz and FWHM of 27 GHz. Also in the literature there are peak frequencies around 390 GHz \cite{bls1}  and frequencies of the order of 356 GHz \cite{bls2}. Similar to the studies using the Raman spectroscopy method, BLS makes it possible to determine a high-frequency peak, which is the result of light scattering by antiferromagnetic magnons, and has a frequency of about 1100 GHz, as shown in Ref.\ \onlinecite{bls3}. Thus, in addition to the determination of the magnon frequencies, we also prove the structural and magnetic quality of our samples. 

\end{document}